\documentclass[fleqn,usenatbib,usedcolumns]{mnras}

\usepackage{amsmath}
\usepackage{txfonts}

\usepackage[british]{babel} 
\usepackage[T1]{fontenc}
\usepackage{ae,aecompl}

\usepackage{graphicx}	
\usepackage{xspace}

\graphicspath{{./Paper_plots/}}
\newcommand{\galform}{{\sc{galform}}\xspace}
\newcommand{\grasil}{{\sc{grasil}}\xspace}

\newcommand{\mum}{$\muup$m}
\newcommand{\hMsol}{$h^{-1}$~M$_{\sun}$}
\newcommand{\hMpc}{$h^{-1}$~Mpc}

\title[Blending bias impacts the halo masses of SMGs]{Blending bias impacts the host halo masses derived from a cross-correlation analysis of bright sub-millimetre galaxies}

\author[W. I. Cowley et al.]{
William I. Cowley$^{1\mathrm{,}2}$\thanks{E-mail: cowley@astro.rug.nl (WIC)},
Cedric G. Lacey$^{1}$,
Carlton M. Baugh$^{1}$,
Shaun Cole$^{1}$,\newauthor
Aaron Wilkinson$^{3}$
\\
$^{1}$Institute for Computational Cosmology, Department of Physics, University of Durham, South Road, Durham, DH1 3LE, UK\\
$^{2}$Kapteyn Astronomical Institute, University of Groningen, PO Box 800, NL-9700 AV Groningen, The Netherlands\\
$^{3}$School of Physics and Astronomy, University of Nottingham, University Park, Nottingham, NG7 2RD, UK
}

\date{Accepted XXX. Received YYY; in original form ZZZ}

\pubyear{2016}

\begin{document}
\label{firstpage}
\pagerange{\pageref{firstpage}--\pageref{lastpage}}
\maketitle

\begin{abstract}
Placing bright sub-millimetre galaxies (SMGs) within the broader context of galaxy formation and evolution requires accurate measurements of their clustering, which can constrain the masses of their host dark matter halos.  Recent work has shown that the clustering measurements of these galaxies may be affected by a `blending bias,' which results in the angular correlation function of the sources extracted from single-dish imaging surveys being boosted relative to that of the underlying galaxies.  This is due to confusion introduced by the coarse angular resolution of the single-dish telescope and could lead to the inferred halo masses being significantly overestimated.  We investigate the extent to which this bias affects the measurement of the correlation function of SMGs when it is derived via a cross-correlation with a more abundant galaxy population.  We find that the blending bias is essentially the same as in the auto-correlation case and conclude that the best way to reduce its effects is to calculate the angular correlation function using SMGs in narrow redshift bins.  Blending bias causes the inferred host halo masses of the SMGs to be overestimated by a factor of $\sim6$ when a redshift interval of $\delta z=3$ is used. However, this reduces to a factor of $\sim2$ for $\delta z=0.5$.  The broadening of photometric redshift probability distributions with increasing redshift can therefore impart a mild halo `downsizing' effect onto the inferred host halo masses, though this trend is not as strong as seen in recent observational studies.     
\end{abstract}




\section{Introduction}
Sub-millimetre galaxies \citep[SMGs, e.g.][]{Blain02,Casey14} are thought to be amongst the most rapidly star-forming objects in the Universe.  They are detected at wavelengths which probe the re-emission of radiation by cold interstellar dust.  Assuming that the initial radiation field is due to star formation\footnote{Studies that have investigated the X-ray properties of SMGs suggest that their bolometric luminosity is dominated by emission from star formation rather than an Active Galactic Nucleus \citep[e.g.][]{Alexander05}.}, the extreme luminosity of this dust leads to prodigious inferred star formation rates of $\gtrsim100$~M$_{\sun}$~yr$^{-1}$ \citep[e.g.][]{Smail02,Swinbank13}.  The shape of a galaxy's spectral energy distribution (SED) at these wavelengths (the Rayleigh-Jeans tail of the dust emission) approximates a power law that decreases with increasing wavelength, meaning that it is subject to a negative $k$-correction \citep[e.g.][]{Blain02}.  For a fixed bolometric luminosity and observer-frame wavelength, shifting the galaxy to higher redshifts means that the SED is sampled at a shorter rest-frame wavelength, where it is intrinsically brighter.  This largely cancels out the effect of dimming due to the increasing luminosity distance, meaning that the observed flux of an SMG is roughly constant over $z\sim1-10$.  Thus SMGs provide a window into (dust-obscured) star formation at high redshift, commonly being found at $z\sim1-3$ \citep[e.g.][]{Chapman05,Simpson13}.

Placing these extreme galaxies into a consistent evolutionary picture remains challenging.  It is not clear what physical mechanisms trigger and quench the star formation rates inferred from observations, and their subsequent evolution to the present day is poorly understood.  Simple arguments which make assumptions about the duration of the extreme star formation event and the subsequent evolution of the stellar populations of the SMGs have been used to argue that they could evolve into massive local elliptical galaxies with most of their stellar mass being assembled during the `SMG phase' \citep[e.g.][]{Swinbank06,Simpson13}, though see Gonz{\'a}lez et al. (\citeyear{Gonzalez11}) for a contrasting view in which this phase accounts for little of the present day stellar mass in their descendants.

A strong constraint on the evolution of a galaxy population can come from observational measurements of its clustering, which provides information regarding the masses of the dark matter halos the galaxies inhabit.  Growth of structure arguments based on results from $N$-body simulations can then be used to infer the distribution of present day host halo mass of the galaxies' descendants \citep[e.g.][]{Fakhouri10}, which can then be compared to the halo masses inferred from the observed clustering of local galaxy populations.  However, the spread in the host halo masses of SMG descendants could be significant ($\sim2$~dex, Cowley et al.~\citeyear{Cowley16}) due to the hierarchical growth of structure.  

Measuring the clustering of FIR/sub-mm galaxies has proven challenging.  Some studies have failed to produce significant detections of clustering \citep[e.g.][]{Scott02,Webb03VI,Coppin06,Williams11}, or the results derived from similar data have proven contradictory \citep[e.g.][]{Cooray10,Maddox10}.  For bright SMGs a significant difficulty is their sparse number density, meaning large area surveys are required to yield sufficient galaxy pairs for the correlation function to be estimated robustly.  An observational study of the clustering of SMGs was performed by \cite{Hickox12}, who ameliorated the problem of small numbers of SMGs by using a cross-correlation \citep{Blake06} with a more abundant \emph{Spitzer} InfraRed Array Camera (IRAC)-selected galaxy population to find that $z=1-3$ SMGs in the LESS\footnote{Large APEX (Atacama Pathfinder EXperiment) Bolometer Camera Array (LABOCA) Extended \emph{Chandra} Deep Field South (ECDFS) Sub-millimetre Survey} source catalogue \citep{Weiss09} have a correlation length of $r_{0}=7.7_{-2.3}^{+1.8}$~\hMpc, corresponding to an inferred halo mass of $M_{\rm halo}=10^{12.8^{+0.3}_{-0.5}}$~{\hMsol}.  This result is consistent with an earlier study by \cite{Blain04} who used measured redshift separations of pairs of SMGs in a number of small fields to estimate a correlation length of $6.9\pm2.1$ $h^{-1}$~Mpc.  Hickox et al. used the median growth rate of haloes from \cite{Fakhouri10} to suggest descendent halo masses consistent with those of local $\sim2-3$~$L_{\star}$ galaxies.

More recently, \cite{Wilkinson16} performed a similar analysis. However, these authors were able to improve upon earlier work by making the first measurements of the clustering of SMGs as a function of redshift, owing to the greater number of SMGs detected as part of the SCUBA-2 (Super Common User Bolometer Array 2, Holland et al. \citeyear{Holland13}) Cosmology Legacy Survey (S2CLS, Geach et al. \citeyear{Geach13,Geach17}) in the UKIDSS--UDS\footnote{United Kingdom Infra-red Telescope (UKIRT) Infra-red Deep Sky Survey -- Ultra Deep Survey} field.  Cross-correlating their SMG sample with a more numerous $K$-band selected galaxy population, Wilkinson et al. estimated that the halo masses of SMGs ranged from $M_{\rm halo}\sim10^{13}$~{\hMsol} at $z\gtrsim2$ to $M_{\rm halo}\sim10^{11}$~{\hMsol} for $1<z<2$.  Wilkinson et al. concluded that the $z\gtrsim2$ SMG population could evolve into local $\sim2-3$~$L_{\star}$ galaxies.

However, the work of Hickox et al. and Wilkinson et al. is based on source catalogues derived from single-dish imaging surveys with a typical angular resolution of $\sim20$~arcsec at full width half maximum (FWHM).  Interferometers such as ALMA (Atacama Large Millimetre Array) have an order of magnitude better resolution, and targeted observations have revealed that many sub-mm sources identified from single-dish imaging are in fact composed of multiple fainter galaxies that could not be distinguished from each other in the original single-dish survey due to its low angular resolution \citep[e.g.][]{Wang11,Hodge13}.  The effect this has on the observed number counts has been investigated \citep{Karim13,Simpson15} but until recently it has been unclear exactly what impact this has on measurements of the clustering of SMGs.

The first predictions for this were made by Cowley et al. (\citeyear{Cowley16}, hereafter C16). There we showed, using the clustering of SMGs predicted using the galaxy formation model of \cite{Lacey16}, that confusion due to the single-dish beam could boost the observed angular correlation function of sub-mm sources by a factor of $\sim4$ relative to the correlation function of the underlying galaxies over all angular scales.  This effect was termed `blending bias'.  Many of the blended sources detected in the simulated imaging of C16 comprise physically unassociated galaxies with a typical redshift separation of $\Delta z\sim1-2$ \citep{Cowley15}.  These galaxies are often fainter than the flux limit of the survey, and are boosted above this by the blending together of their flux by the beam. Their positions would not be included in the source catalogue otherwise.  Though the galaxies that have their flux blended into a single source are generally chance projections along the line of sight, their positions are correlated with the positions of other galaxies at the same redshift.  Some of these will also be included in the source catalogue, which leads to `beam-induced'  correlated pairs of sources resulting in a boost in the correlation function (blending bias) on all angular scales.  Furthermore, C16 showed that the redshift interval considered has an impact on the blending bias, with narrower redshift intervals including fewer of these beam-induced pairs and so resulting in much smaller blending bias factors.

Wilkinson et al. performed an analysis similar to C16 for the redshift intervals considered in their work ($\Delta z=0.5$ for $1.0<z<3.5$) and found a blending bias factor of $\sim1.2$ independent of redshift.  However, this was for the auto-correlation of sub-mm sources, and also did not consider the effect that the broadening with increasing redshift of the photometric redshift probability distributions of their galaxies would have on the blending bias.  Here we present predictions for the blending bias when the correlation function of sub-mm sources is determined via a cross-correlation with a more abundant galaxy population.  Also, in order to provide the best possible comparison of the observations of Wilkinson et al. and the galaxy formation model used in C16, we choose a $K$-band sample of similar depth and use the same redshift intervals considered in Wilkinson et al.  We also mimic, to first order, the effect of broader photometric redshift probability distributions with increasing redshift on the clustering measurement.  In addition, the nature of our simulations allows us to make predictions for the field-to-field variation expected for such observations.  In this paper we focus on observations and predictions made at $850$~$\muup$m, however we also expect that blending bias will affect clustering analyses made at shorter FIR wavelengths with \emph{Herschel}-SPIRE\footnote{Spectral and Photometric Imaging Receiver} \citep[e.g.][]{Cooray10,Maddox10}, where confusion is also significant \citep[e.g.][]{Nguyen10}.    

This paper is structured as follows: In Section \ref{sec:Model} we briefly describe the galaxy formation model, the model for computing the dust emission of the simulated galaxies at sub-mm wavelengths and the method for creating the simulated imaging. In Section \ref{sec:Results} we present our main results\footnote{Some of the model data presented here will be made available at \url{http://icc.dur.ac.uk/data/}.  For other requests please contact the first author.} and we conclude in Section \ref{sec:conclusion}.  Throughout we assume a flat $\Lambda$ cold dark matter ($\Lambda$CDM) cosmology with cosmological parameters consistent with the $7$~year \emph{Wilkinson Microwave Anisotropy Probe} (\emph{WMAP7}) results \citep{Komatsu11} i.e. $\Omega_{\rm m}=0.272$, $\Omega_{\rm b}=0.0455$, $\Omega_{\Lambda}=0.728$, $h=0.704$, $\sigma_{8}=0.81$, $n_{\rm s}=0.967$.

\section{The Theoretical Model}  
\label{sec:Model}
Here we introduce our model, which combines a dark matter only $N$-body simulation, a state-of-the-art semi-analytic model of galaxy formation and a simple model for the reprocessing of stellar radiation by dust in which the dust temperature is calculated self-consistently based on radiative transfer and global energy balance arguments.  For further details we refer the reader to \cite{Lacey16}.  We also briefly describe our method for creating the simulated imaging used throughout. 
\subsection{GALFORM}
The Durham semi-analytic model of hierarchical galaxy formation, \galform, was introduced in \cite{Cole00}, building on ideas outlined by \cite{WhiteRees78}, \cite{WhiteFrenk91} and \cite{Cole94}.  Galaxy formation is modelled \emph{ab initio}, beginning with a specified cosmology and a linear power spectrum of density fluctuations and ending with predicted galaxy properties at different redshifts.

Galaxies are assumed to form from baryonic condensation within the potential wells of dark matter halos with their subsequent evolution being controlled in part by the merging history of the halo. Here, the halo merger trees are extracted directly from a dark matter only $N$-body simulation \citep[e.g.][]{Helly03,Jiang14}. We use a ($500$~\hMpc)$^{3}$ Millennium-style simulation \citep{Springel05,Guo13} with cosmological parameters consistent with \emph{WMAP7} results \citep{Komatsu11}, hereafter referred to as MR7.  Halo masses are as defined using the DHalo algorithm \citep{Jiang14}.

The baryonic processes thought to be important for galaxy formation are included as a set of coupled differential equations which essentially track the exchange of mass and metals between stellar, cold disc gas and hot halo gas components in each galaxy.  Stellar luminosities are computed through coupling the resulting star formation and metal enrichment histories of the simulated galaxies with evolutionary population synthesis models \citep[e.g.][]{Maraston05}.  The values of the parameters in these simplified equations which describe the complex physical processes involved are then calibrated against a predetermined set of data from observations and simulations, which provides a strong constraint on the available parameter space \citep[e.g.][]{Lacey16}.  In particular, the \cite{Lacey16} model is calibrated to reproduce the observed optical and near infra-red luminosity functions for $z\lesssim3$ and, importantly for this work, the sub-mm galaxy number counts at $850$~{\mum}. 

In this model SMGs occupy halos in the mass range $M_{\rm halo}\sim10^{11.5}-10^{12}$~{\hMsol} over a large range of redshifts ($0.2\lesssim z\lesssim4$). This is because the interplay of physical processes such as gas cooling, supernova feedback and radio-mode AGN feedback means this represents the halo mass range most conducive to star formation in the model (Lacey et al. \citeyear{Lacey16}, C16).        

\subsection{The Dust Model}
To determine a simulated galaxy's far infra-red (FIR) flux, a model is required to calculate the absorption and re-emission of stellar radiation by interstellar dust.    

We assume that interstellar dust exists in two components, spherical molecular clouds of a fixed gas surface density in which stars form and a diffuse interstellar medium distributed smoothly throughout an exponential disc.  The energy from stellar radiation absorbed by each component can be calculated using the star formation and metal enrichment history for a galaxy predicted by \galform, and then solving the equations of radiative transfer in this assumed geometry.  The dust emission is then calculated using global energy balance arguments and assuming that it emits as a modified blackbody.  Crucially, this means that dust temperature is not a free parameter in the model but is calculated self-consistently for each dust component in each galaxy.

Despite its simplicity the model is able to reproduce the predictions of the more sophisticated spectrophotometric code \grasil \citep{Silva98}, offset only by minor factors of $\lesssim2$ with very a tight scatter, for $\lambda_{\rm rest}\gtrsim70$~{\mum} \citep{CowleyDustSEDs16}.  The sheer computational expense of codes such as \grasil ($\sim3-5$ CPU mins per galaxy) however, makes them unsuitable for the large number of galaxies contained in the cosmological volumes used in this work.

\subsection{Creating Simulated Imaging of SMGs}
\begin{figure}
\centering
\includegraphics[width = \columnwidth]{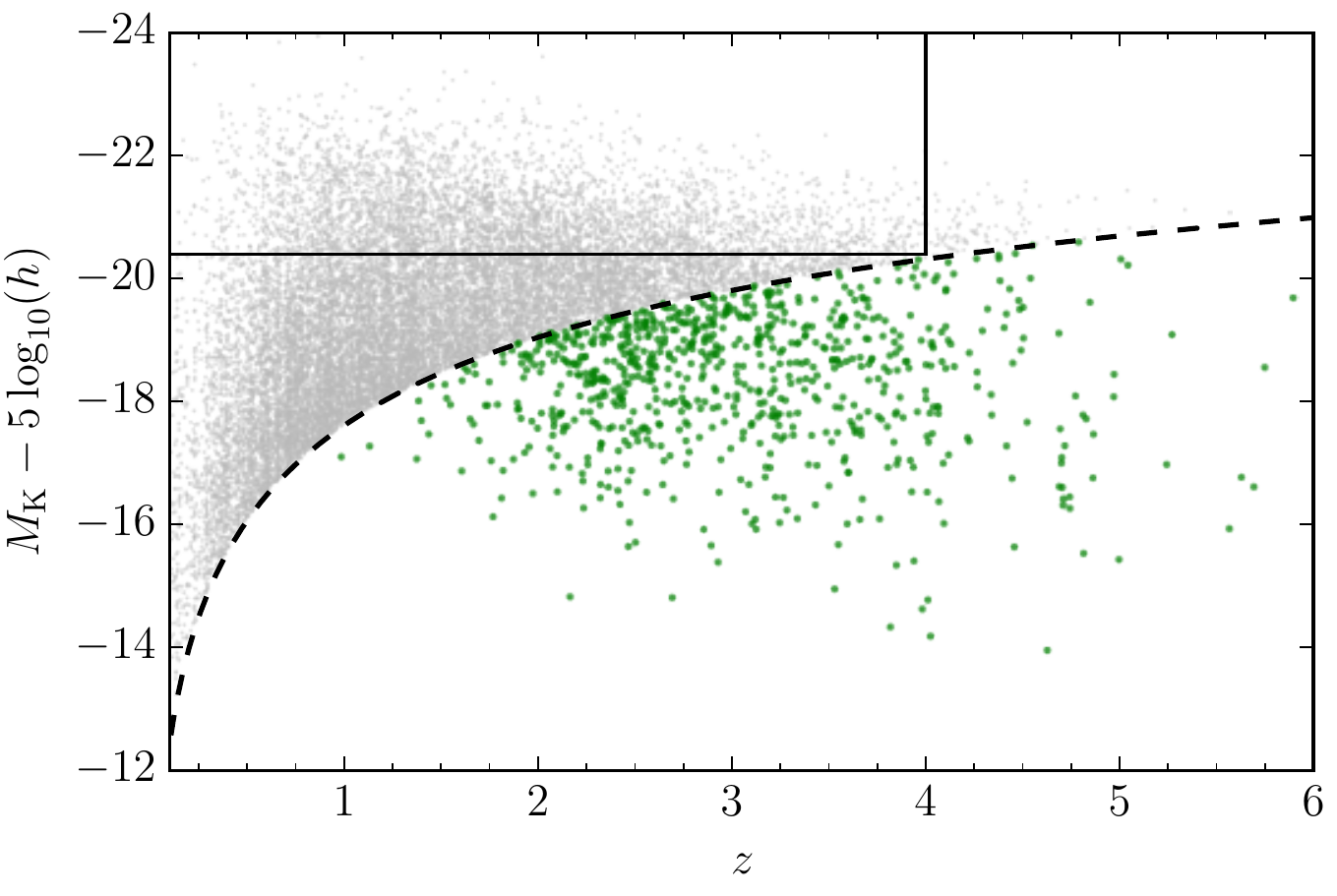}
\caption{The $K$-band absolute magnitude versus redshift for lightcone galaxies flux limited in the $K$-band at an apparent magnitude of $25$ (grey dots) for one of the $50\times4$~deg$^2$ fields. For clarity, only $1$~per~cent of this sample is shown.  The $K$-band flux limit is indicated by the dashed black line.  The green points indicate galaxies with $S_{850\muup\mathrm{m}}>4$~mJy that are not selected in the $K$-band.  The black solid lines indicate a volume-limited $K$-band sample for $z<4$.  All magnitudes are in the AB system.}
\label{fig:absK_z}
\end{figure}
\begin{figure}
\centering
\includegraphics[width = \columnwidth]{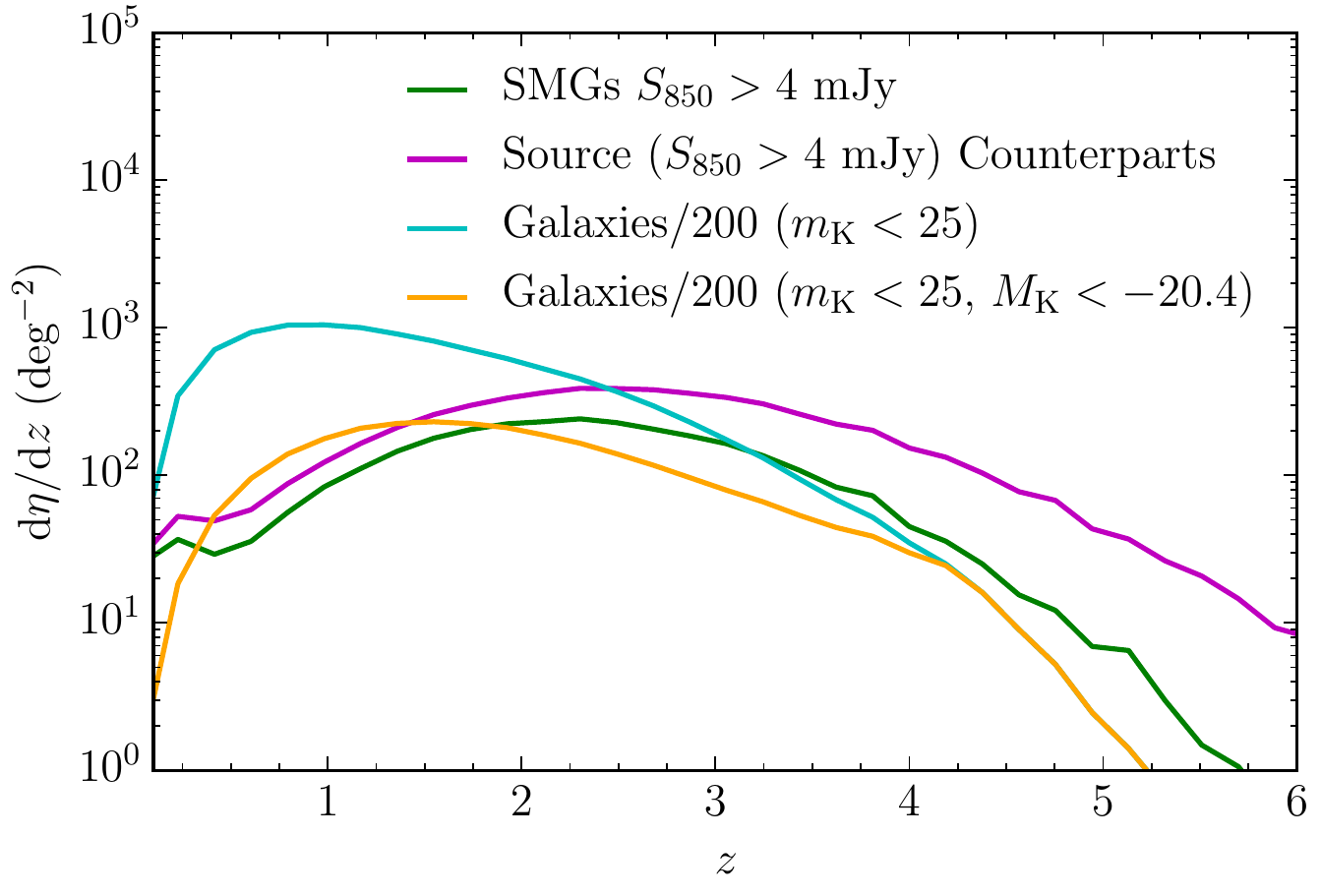}
\caption{Predicted average redshift distributions from our $50\times4$~deg$^2$ fields for $S_{850\muup\rm m}>4$~mJy galaxies (green line), the counterparts (see text) of sources with $S_{850\muup\mathrm{m}}>4$~mJy extracted from the simulated sub-mm imaging (magenta line), flux-limited sample of $K$-band selected galaxies (cyan line) and a volume-limited sample (for $z<4$) of $K$-band selected galaxies (orange line).  The latter two lines have both been divided by a factor of $200$ for presentation purposes.  All magnitudes are in the AB system.}  
\label{fig:dndz}
\end{figure}
The lightcone code presented in \cite{Merson13} is used to create mock surveys of SMGs using $50$ randomly orientated lines of sight through the simulation volume.  For the purposes of this study we use an AB apparent magnitude of $m_{\rm K}<25$ to select our $K$-band population, similar to that used by \cite{Wilkinson16}, and a limit of $S_{850\muup\rm m}>0.35$~mJy.  This $850$~{\mum} limit is chosen as it is the flux above which $90$ per cent of the total predicted background light is included in a typical image.  This prediction is in good agreement with the observations of \cite{Puget96} and \cite{Fixsen98}, and so means that our sub-mm maps have a realistic extragalactic background light \citep{Cowley15}.  We choose a maximum area of $4$~deg$^{2}$ for our fields, which is larger than currently surveyed in sub-mm observations, to reduce the effect of a finite survey area on our results.  An example of the $K$-band absolute magnitude versus redshift for the resulting input catalogue is shown in Fig.~\ref{fig:absK_z}.

Following the method presented in \cite{Cowley15}, galaxies in each mock catalogue are mapped onto a grid of pixels according to their position, such that the value in a pixel is equal to the sum of the $850$~{\mum} flux of all the galaxies that fall within it.  The resulting image is then smoothed with a Gaussian beam, with the pixel scale chosen such that the beam is well sampled.  Instrumental (white) noise is then added and matched-filtering is performed prior to source extraction.  In order to mimic the SCUBA-2 observational data we choose a Gaussian with a $15$ arcsec FWHM and $\sim1$~mJy/beam of instrument noise.  Source counterparts are identified as the galaxy which makes the dominant contribution to the overall sub-mm flux of the source.  The statistics of the resulting source catalogue can then be compared to those of the underlying galaxies.  The resulting redshift distributions are compared in Fig.~\ref{fig:dndz}.  We can see that the source counterparts are more numerous than galaxies at the same flux limit, and that their distribution has a more prominent high-redshift tail.  The surface number densities of the flux-limited $K$-band sample, the $S_{850\muup\rm m}>4$~mJy galaxies and the counterparts to ${S_{850\muup\rm m}>4}$~mJy sources are $4.02\times10^{5}$, $5.54\times10^{2}$ and $1.05\times10^{3}$~deg$^{-2}$, respectively. 

\section{Results}
\label{sec:Results}
The simplest measure of clustering from a galaxy imaging survey is the two-point angular correlation function $w(\theta)$. The probability of finding two objects separated by an angle $\theta>0$ is defined as \citep[e.g.][]{Peebles80}
\begin{equation}
\delta P_{12}(\theta) = \eta^{2}\,[1+w(\theta)]\,\delta\Omega_{1}\delta\Omega_{2}\rm,
\end{equation} where $\eta$ is the mean surface density of objects per unit solid angle and $\delta\Omega_{\rm i}$ is a solid angle element, such that $w(\theta)$ represents the excess probability of finding objects at angular separation, $\theta$, relative to a random (Poisson) distribution.  The measured angular correlation function for a given galaxy population, $w_{\rm g}$, can then be compared to that expected for the dark matter, $w_{\rm DM}$, to yield the large-scale bias of the galaxy population, calculated as
\begin{equation}
b_{\rm g}(\theta) = \left[\frac{w_{\rm g}(\theta)}{w_{\rm DM}(\theta)}\right]^{1/2}\rm.
\end{equation}  
Although bias is scale dependent \citep[e.g.][]{Angulo08} it is usually approximated as a constant on large (linear) scales, where it can be compared to an expected bias computed by a weighted average of the bias values over the halos that are occupied \citep[e.g.][]{CooraySheth02} 
\begin{equation}
b_{\rm eff}(z) = \frac{\int b(M_{\rm h},z)\,n(M_{\rm h},z)\,\langle N_{\rm gal}|M_{\rm h}\rangle\,\mathrm{d}M_{\rm h}}{\int n(M_{\rm h},z)\,\langle N_{\rm gal}|M_{\rm h}\rangle\,\mathrm{d}M_{\rm h}}\rm.
\label{eq:bias_eff}
\end{equation}  Here, $b(M_{\rm h},z)$ is the large-scale bias of halos with mass $M_{\rm h}$ at redshift $z$, $n(M_{\rm h},z)$ is the halo mass function at redshift $z$ such that $n(M_{\rm h},z)\,\mathrm{d}M_{\rm h}$ is the comoving number density of halos in the mass range $[M_{\rm h},M_{\rm h}+\mathrm{d}M_{\rm h}]$, and $\langle N_{\rm gal}|M_{\rm h}\rangle$ is the mean of the halo occupation distribution (HOD, the expected mean number of galaxies within a halo of mass $M_{\rm h}$).  Thus measuring the large-scale bias can yield information regarding the halo masses that the galaxy population occupy.
   
\subsection{A cross-correlation analysis of sub-mm sources}
\label{subsec:xcorr_smgs}
\begin{figure}
\centering
\includegraphics[width = \columnwidth]{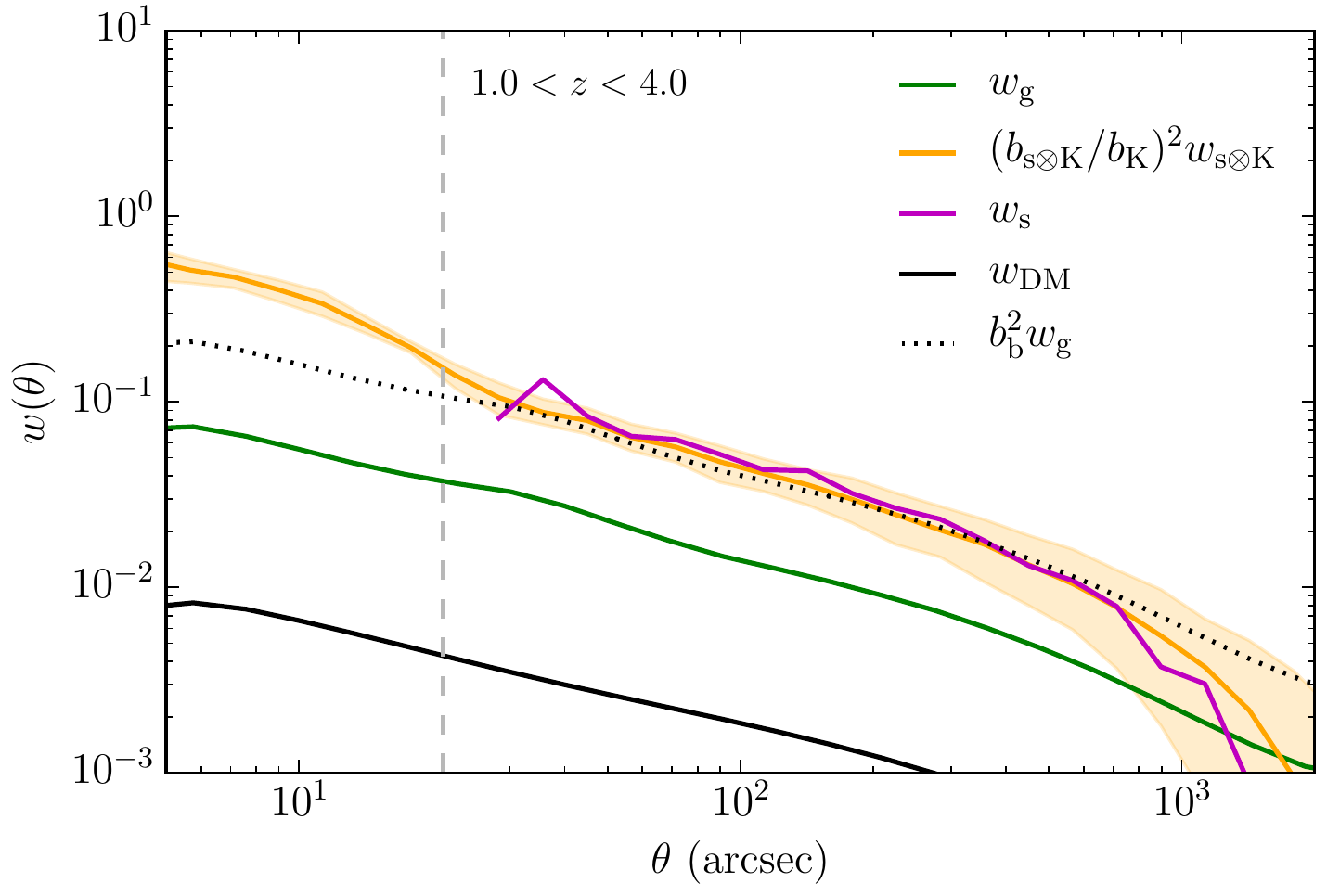}
\caption{Predicted angular correlation functions in the redshift range $1.0<z<4.0$.  The angular correlation function of galaxies selected by $S_{850\muup\rm m}>4$~mJy is shown by the green line.  The cross-correlation of counterparts to sources with $S_{850\muup\rm m}>4$~mJy with a volume-limited $K$-band selected sample, averaged over our $50\times4$~deg$^2$ fields and scaled to remove the bias of the $K$-band sample is shown by the orange line. The shaded orange region corresponds to the predicted $1\sigma$ ($16-84$ percentile) field-to-field variation for the $4$~deg$^{2}$ field area used. The auto-correlation of the source counterparts (averaged over $50\times4$~deg$^{2}$) is shown by the magenta line, the correlation function of dark matter in the MR7 simulation is shown by the black line, and the correlation function of the galaxies scaled by the blending bias squared (here $b_{\rm b}=1.7$) is shown by the black dotted line.  The vertical dashed line indicates the FWHM of the match-filtered point spread function used to create the simulated imaging $\sim21.2$ arcsec.}
\label{fig:wtheta_1p0z4p0}
\end{figure}
\begin{figure}
\centering
\includegraphics[width=\columnwidth]{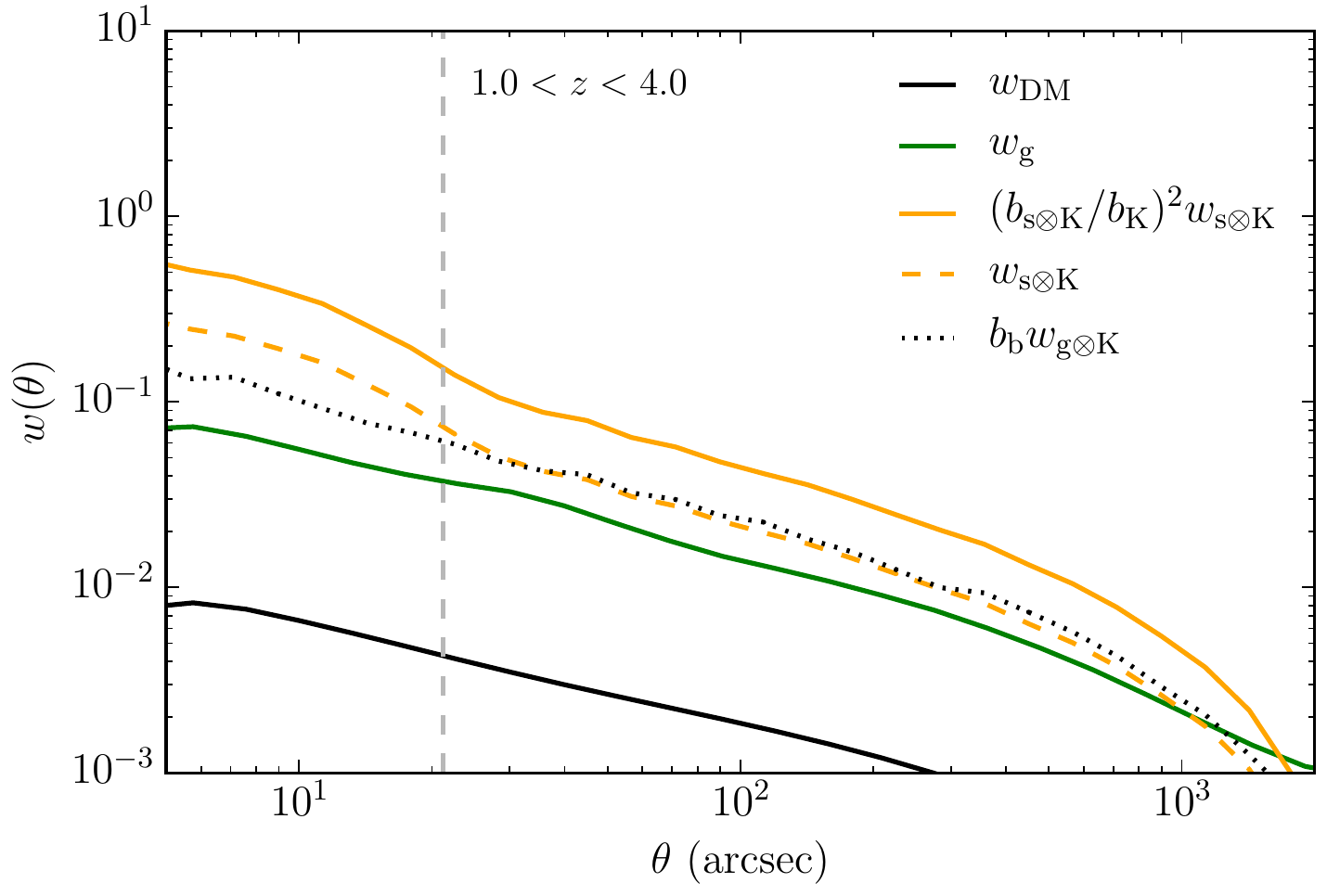}
\caption{Predicted angular correlation functions in the redshift range $1.0<z<4.0$.  The dashed orange line shows the cross-correlation of counterparts to sources with $S_{850\muup\rm m}>4$~mJy with a volume-limited $K$-band selected sample, averaged over our $50\times4$~deg$^2$ fields.  The dashed black line shows the cross-correlation of galaxies with $S_{850\muup\rm m}>4$~mJy with a volume-limited $K$-band selected sample, averaged over $50\times4$~deg$^2$ fields, and scaled by the blending bias.  All other lines have the same meaning as in Fig.~\ref{fig:wtheta_1p0z4p0}.}
\label{fig:wtheta_xcorr_1p0z4p0}
\end{figure}
\begin{figure*}
\includegraphics[width = \linewidth]{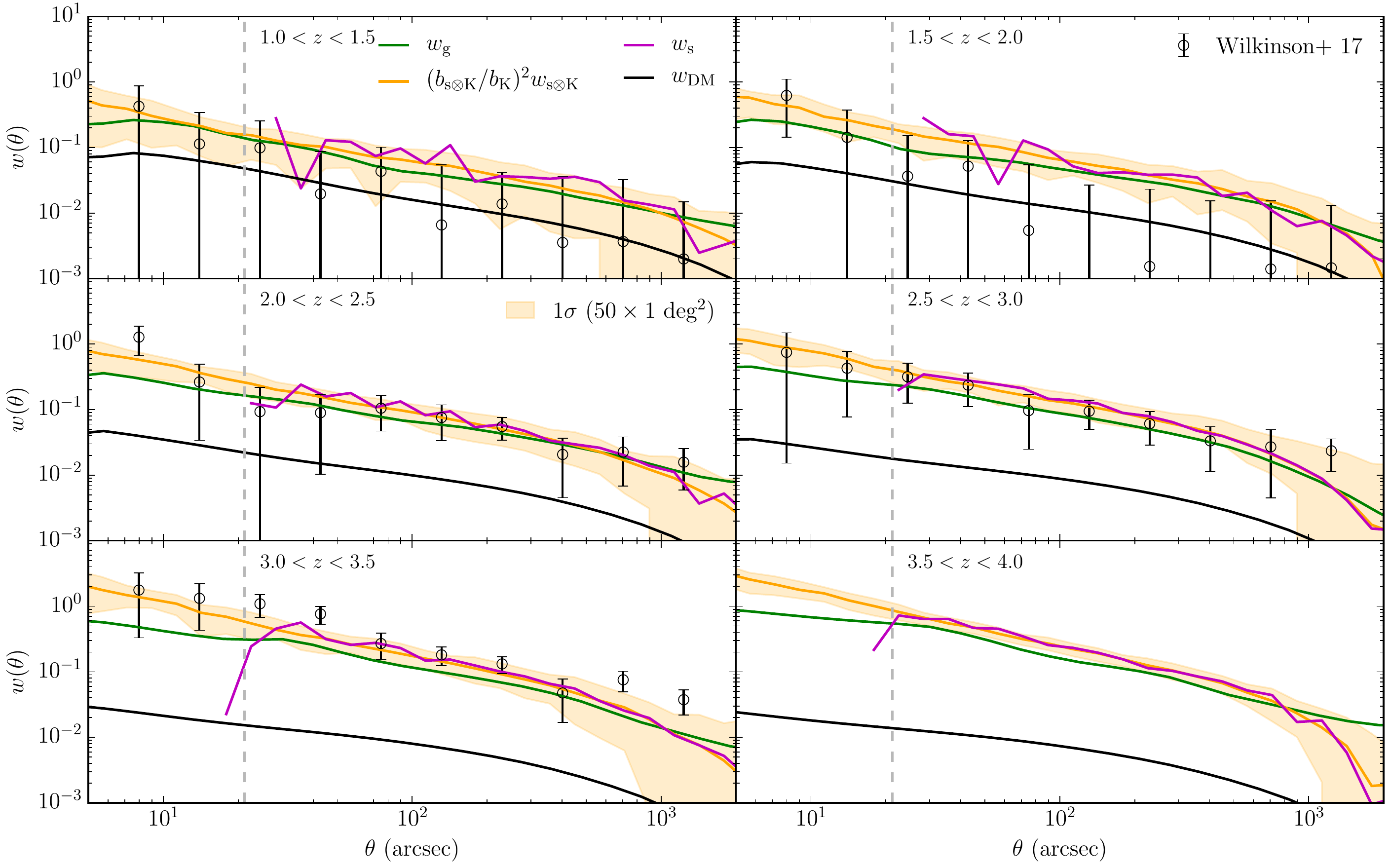}
\caption{Predicted angular correlation functions for different redshift intervals indicated in the panels for galaxies selected with $S_{850\muup\rm m}>4$~mJy (green lines), the cross-correlation of counterparts to sources with $S_{850\muup\rm m}>4$~mJy with a volume-limited $K$-band selected sample, averaged over $50\times4$~deg$^{2}$ fields and scaled so as to remove the bias of the $K$-band sample (orange line), and the auto-correlation of the source counterparts (averaged over the $50\times4$~deg$^{2}$ fields, magenta line).  We show also observational data from Wilkinson et al. (\citeyear{Wilkinson16}), which are derived from a cross-correlation of sources with a $K$-band selected galaxy sample, and so should be compared with our orange line.  The shaded orange region corresponds to the $1\sigma$ ($16-84$ percentile) scatter derived from field-to-field variations, calculated from the central $1$~deg$^{2}$ region in each of our fields in order to match the area used in the observations of Wilkinson et al.  The vertical dashed line indicates the FWHM of the match-filtered point spread function used to create the simulated imaging $\sim21.2$ arcsec.}
\label{fig:wtheta_obs_W16}
\end{figure*}
\begin{figure}
\centering
\includegraphics[width = \columnwidth]{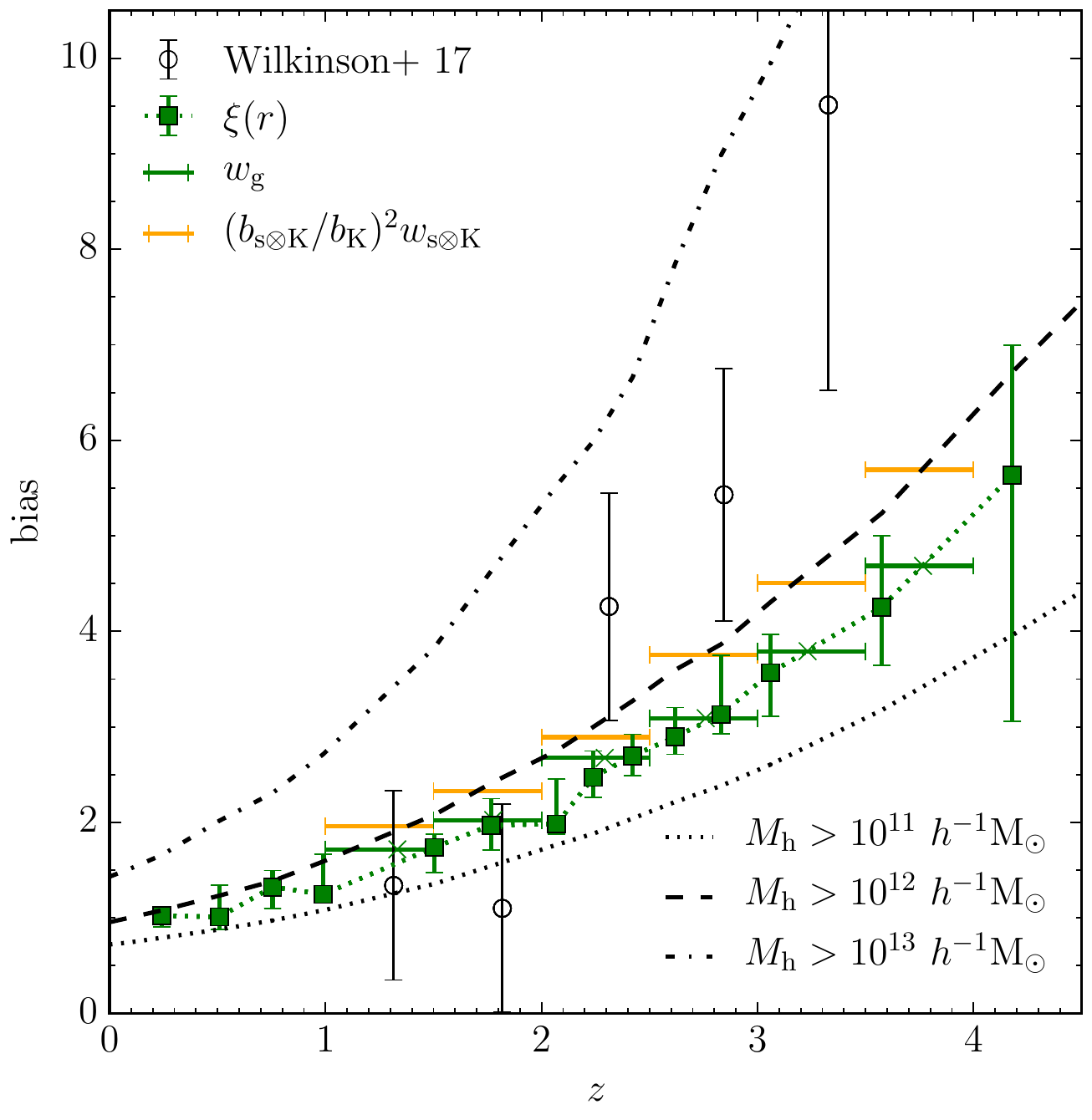}
\caption{Predicted evolution of large-scale bias with redshift.  Green squares with errorbars represent the bias measured directly from the 3D spatial correlation function of SMGs with $S_{850\muup\rm m}>4$~mJy, as is done in Cowley et al. (\citeyear{Cowley16}).  The $1\sigma$ errors are calculated using the volume bootstrap method advocated in Norberg et al. (\citeyear{Norberg09}). The horizontal green bars show the large-scale bias of the SMGs with $S_{850\muup\rm m}>4$~mJy derived from the angular correlation function over the redshift range indicated by the width of the bar.  The horizontal orange bars show the same but for the angular correlation function of sub-mm sources calculated via a cross-correlation with a volume-limited $K$-band selected sample.  The dotted, dashed and dash-dotted black lines show the evolution of the large-scale bias of halos with $M_{\rm halo}>10^{11}$, $10^{12}$ and $10^{13}$~{\hMsol} respectively, measured directly from the MR7 simulation.  Observational data (black circles with errors) are from Wilkinson et al. (\citeyear{Wilkinson16}).}
\label{fig:bias_evol_W16}
\end{figure}

As we are comparing the predictions of our model to the analysis of \cite{Wilkinson16}, we begin with the source catalogue ($S_{850~\muup\rm m}>4$~mJy) derived from source extraction from the simulated images of C16 as described previously.  The SMG sample used by Wilkinson et al. has a slightly fainter flux limit ($\sim3.5$~mJy, Chen et al. \citeyear{TC16}), however we do not expect this to have a significant impact on our science results.  In C16 we showed that the angular auto-correlation of the sub-mm sources, $w_{\rm s}$, was boosted by a `blending bias' factor, $b_{\rm b}$, relative to that of the underlying galaxy population, $w_{\rm g}$, such that $w_{\rm s}=b_{\rm b}^{2}w_{\rm g}$.  In this paper we calculate $w_{\rm s}$ via a cross-correlation with a volume-limited $K$-band selected galaxy population ($m_{\rm K}<25$).  Assuming linear theory, the large-scale bias of the sub-mm sources, $b_{\rm s}$, can be determined using
\begin{equation}
b_{\rm s} = b_{\rm s\otimes\rm K}^{2}/b_{K}\rm ,
\end{equation}
where $b_{\rm K}$ represents the bias of the $K$-band selected galaxy population as measured from its auto-correlation function and $b_{\rm s\otimes\rm K}$ is the bias of the cross-correlation of the two populations.  This means that $(b_{\rm s\otimes\rm K}/b_{\rm K})^{2}w_{\rm s\otimes K}$ is equivalent to $w_{\rm s}$, provided that the blending bias effects both measurements in the same way.
 
To calculate the angular cross-correlation of the sub-mm sources and the $K$-band galaxy sample, $w_{\rm s\otimes\rm K}$, we use the \cite{LandySzalay93} estimator adapted for cross-correlations
\begin{equation}
w_{\rm s\otimes\rm K}(\theta) = \frac{DD_{\rm sK}-DR_{\rm sK}-DR_{\rm Ks}+RR_{\rm sK}}{RR_{\rm sK}}\rm,
\end{equation}
where $DD$, $DR$ and $RR$ represent data-data, data-random and random-random pairs respectively, and the subscripts $\rm{s}$ and $\rm{K}$ represent the sub-mm sources and $K$-band selected galaxies respectively.  In calculating $w_{\rm s\otimes\rm K}$ we use the actual number of sources in each field to estimate the mean surface density, rather than the true surface density.  This causes the angular correlation function to be underestimated by an average amount, $\sigma^{2}$, often referred to as the integral constraint \citep{GrothPeebles77}.  For the cross-correlation functions this quantity is related to the field-to-field variation in the number counts through
\begin{equation}
\sigma_{\rm s\otimes K}^{2}=\frac{\langle(\eta_{\rm s}-\langle\eta_{\rm s}\rangle)(\eta_{\rm K}-\langle\eta_{\rm K}\rangle)\rangle}{\langle\eta_{\rm s}\rangle\langle\eta_{\rm K}\rangle} - \frac{\langle\eta_{\rm sK}\rangle}{\langle\eta_{\rm s}\rangle\langle\eta_{\rm K}\rangle}\rm,
\label{eq:sigma_Xcorr}
\end{equation}
where $\eta_{\rm sK}$ represents the surface density of objects that are in both populations. We evaluate this quantity for our $50$ lightcone fields and add it onto our computed cross-correlation functions.  We also make the corresponding correction to our auto-correlation functions.  These corrections are typically on the order of $\sim10^{-3}$.  We note that equation (\ref{eq:sigma_Xcorr}) is not how this correction is usually calculated in observational studies, where the expression $\sigma^{2}=\sum RR(\theta)w(\theta)/\sum RR(\theta)$ is more commonly used to evaluate the integral constraint, in the absence of multiple fields. However we have checked that this expression gives essentially identical results to equation~(\ref{eq:sigma_Xcorr}).

In Fig~\ref{fig:wtheta_1p0z4p0} we show the angular cross-correlation function of sub-mm sources with the $K$-band galaxy population, and (for comparison) the auto-correlation of sub-mm sources, over the redshift range $1<z<4$.  For our sub-mm sources we use the position and redshift of the galaxy that makes the largest contribution to the flux of the source.  The angular correlation functions for the galaxies and dark matter are calculated from their spatial correlation functions using the \cite{Limber53} equation (computed using a method similar to that described in Gonzalez-Perez et al. \citeyear{GonzalezPerez_ERG}), appropriately changing the redshift limits, as is done in C16\footnote{In principle these could be derived from lightcone catalogues giving essentially identical results, however we prefer using Limber's equation as it utilises all of the clustering information in our simulation volume.}.  We derive a blending bias factor of $b_{\rm b}\sim1.7$ comparing the clustering of sub-mm sources and galaxies. For reference we also show the galaxy correlation function scaled by $b_{\rm b}^{2}$.  For calculating the biases we restrict ourselves to the angular range over which the dark matter correlation function is approximately linear.  We do this by excluding scales for which $w_{\rm{DM,non-linear}}>1.2\times w_{\rm{DM,linear}}$ from our computation of the bias.  We also exclude angular scales larger than $10^3$~arcsec to ensure that the bias measurements are not affected by the finite area of our mock surveys.  We can see that the auto-correlation and the scaled cross-correlation functions are essentially the same.  It therefore appears that blending bias behaves in a similar manner to a linear scale-independent bias.  In this regime the ratio of the cross-correlation of the $K$-band sample with the sub-mm sources, to the cross-correlation of the $K$-band sample with sub-mm galaxies, should simply be equal to the blending bias i.e. $w_{\rm s\otimes K}=b_{\rm b}w_{\rm g\otimes K}$.  We show that this is the case in Fig.~\ref{fig:wtheta_xcorr_1p0z4p0}.  

Thus, whilst the cross-correlation technique can provide smaller statistical errors than the auto-correlation due to the larger number of objects considered, it is still affected by blending bias in the same way.

In order to compare the predictions of our model to the observations of \cite{Wilkinson16} we repeat this analysis using their quoted redshift intervals with $\Delta z=0.5$.  This is shown in Fig.~\ref{fig:wtheta_obs_W16}. We also show the predicted $16-84$ percentile field-to-field variance, estimated from $50$ lightcone fields.  For calculating the predicted field-to-field variance we assume an area of $1$~deg$^{2}$, comparable to that used in Wilkinson et al.  

The agreement between the model and the observations appears to be generally favourable, with the majority of observed data points in each redshift bin (apart from the $1.5<z<2$ bin) lying within the predicted $1\sigma$ region, indicating that the model is broadly consistent with the observed data.    

We can also see from Fig.~\ref{fig:wtheta_obs_W16} the blending bias factors have been reduced (to $b_{\rm b}\sim1.1-1.2$) due to the narrower redshift interval than considered previously. Again, they are essentially the same as those that would be derived from the auto-correlation of the sub-mm sources and are very similar to those derived in \cite{Wilkinson16} for the auto-correlation case (see their Table 2). 
 
In Fig.~\ref{fig:bias_evol_W16} we show the large-scale bias calculated from the cross-correlation derived function, compared to that of the actual underlying galaxies.  We can see that blending bias still affects the inferred halo mass of the SMGs, although to a much lesser extent than it would for the broader $1<z<4$ redshift interval, where $b_{\rm b}\sim1.7$.  Using the large-scale bias - halo mass relations of \cite{SMT01} we find that the blending bias ($b_{\rm b}\sim1.1-1.2$) results in the halo masses of SMGs being overestimated by a factor of $\sim2$.  For the broader $1<z<4$ redshift interval this overestimate is a factor of $\sim6$.  For this we have assumed that all galaxies occupy host dark matter halos of the same mass [i.e. the $\langle N_{\rm gal}|M_{\rm h}\rangle$ term in equation~(\ref{eq:bias_eff}) is described by a Dirac delta function] and used the median redshift of the relevant population (sub-mm galaxies or sub-mm source counterparts) in the redshift interval considered.  We also show in Fig~\ref{fig:bias_evol_W16}, for comparison, the large-scale bias values derived by Wilkinson et al. (\citeyear{Wilkinson16}), though recomputed assuming the same \emph{WMAP7} cosmological parameters as assumed in this work. 

Immediately apparent from Fig~\ref{fig:bias_evol_W16} is that despite the general agreement between the predicted and observed correlation functions seen in Fig.~\ref{fig:wtheta_obs_W16} the inferred large-scale bias values do not agree.  We attribute this to photometric redshift probability distributions used for the observed galaxies, and discuss this in more detail in the next Section.  We list our results from this section, the predicted large-scale sub-mm galaxy bias ($b_{\rm g}$), blending bias ($b_{\rm b}$) and large-scale sub-mm source bias ($b_{\rm s}$) for each $\Delta z=0.5$ redshift interval [listed in column (a)] in Table ~\ref{table:results}.  The table also lists the results from Section~\ref{subsec:photzs}, where we use the redshift intervals described in column (b), as discussed below.  For reference we also list the large-scale bias values derived by \cite{Wilkinson16}.
 
\begin{table*}
\centering
\caption{Predicted large-scale bias of sub-mm galaxies ($b_{\rm g}$), blending bias ($b_{\rm b}$) and large-scale bias of sub-mm sources ($b_{\rm s}$, note that $b_{\rm g}b_{\rm b}=b_{\rm s}$) for the top-red redshift intervals indicated in columns (a) and (b).  The large-scale bias observed by Wilkinson et al. is also shown.}
\begin{tabular}{ccccccccc}\hline
(a)&$b_{\rm g}$&$b_{\rm b}$&$b_{\rm s}$&(b)&$b_{\rm g}$&$b_{\rm b}$&$b_{\rm s}$&$b_{\rm s}$\\
& & & & & & & &(Wilkinson et al.)\\\hline
$1.0<z<1.5$&$1.7$&$1.1$&$2.0$&$0.9<z<1.6$&$1.7$&$1.1$&$1.9$&$1.34\pm0.99$\\
$1.5<z<2.0$&$2.0$&$1.1$&$2.3$&$1.3<z<2.2$&$2.1$&$1.2$&$2.5$&$1.10\pm1.09$\\
$2.0<z<2.5$&$2.7$&$1.1$&$2.9$&$1.7<z<2.8$&$2.6$&$1.3$&$3.3$&$4.26\pm1.19$\\
$2.5<z<3.0$&$3.1$&$1.2$&$3.8$&$2.1<z<3.3$&$3.1$&$1.4$&$4.2$&$5.43\pm1.32$\\
$3.0<z<3.5$&$3.8$&$1.2$&$4.5$&$2.3<z<4.2$&$3.4$&$1.5$&$5.0$&$9.51\pm2.99$\\\hline
\multicolumn{9}{l}{(a) Top-hat redshift interval used in Section~\ref{subsec:xcorr_smgs} and quoted by Wilkinson et al.}\\
\multicolumn{9}{p{0.6\linewidth}}{(b) Top-hat redshift interval used in Section~\ref{subsec:photzs}, chosen such that the normalisation of the dark matter correlation function is the same as used by Wilkinson et al.}\\
\hline 
\end{tabular}
\label{table:results}
\end{table*}   

\subsection{The effect of photometric redshifts}
\label{subsec:photzs}
\begin{figure}
\centering
\includegraphics[width = \columnwidth]{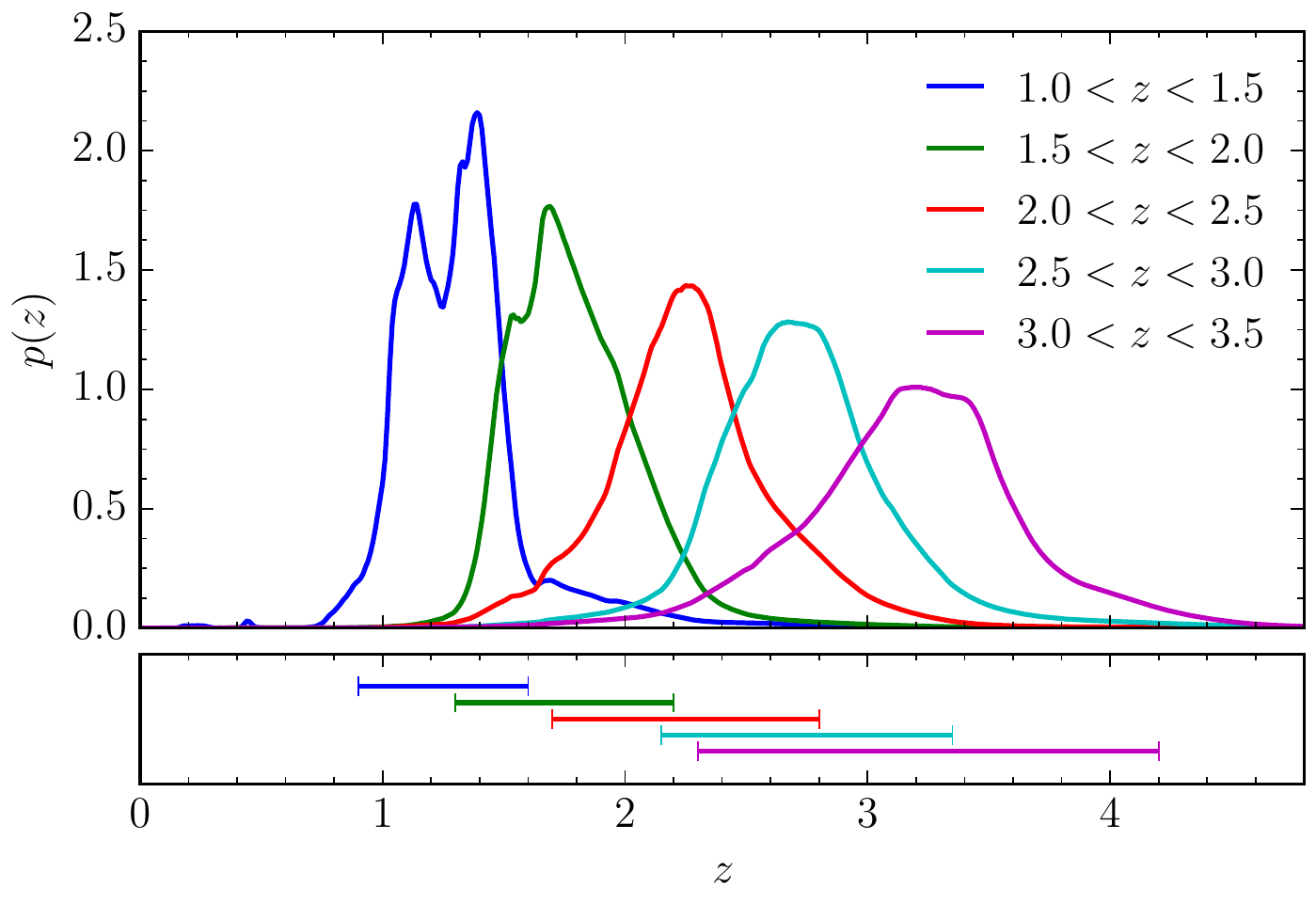}
\caption{\emph{Top panel:} Sub-millimetre galaxy photometric redshift distributions from Wilkinson et al. (\citeyear{Wilkinson16}).  The distributions are shown for the redshift intervals indicated in the legend and are normalised to have unit area.  \emph{Bottom panel:} The width of the top-hat redshift interval required (with the same central redshift) so that the angular dark matter correlation functions computed using the predicted redshift distributions in Fig.~\ref{fig:dndz} have the same normalisation as those computed using the redshift distributions in the top panel.}
\label{fig:Wilkinson16_pzs}
\end{figure}
\begin{figure}
\centering
\includegraphics[width=\columnwidth]{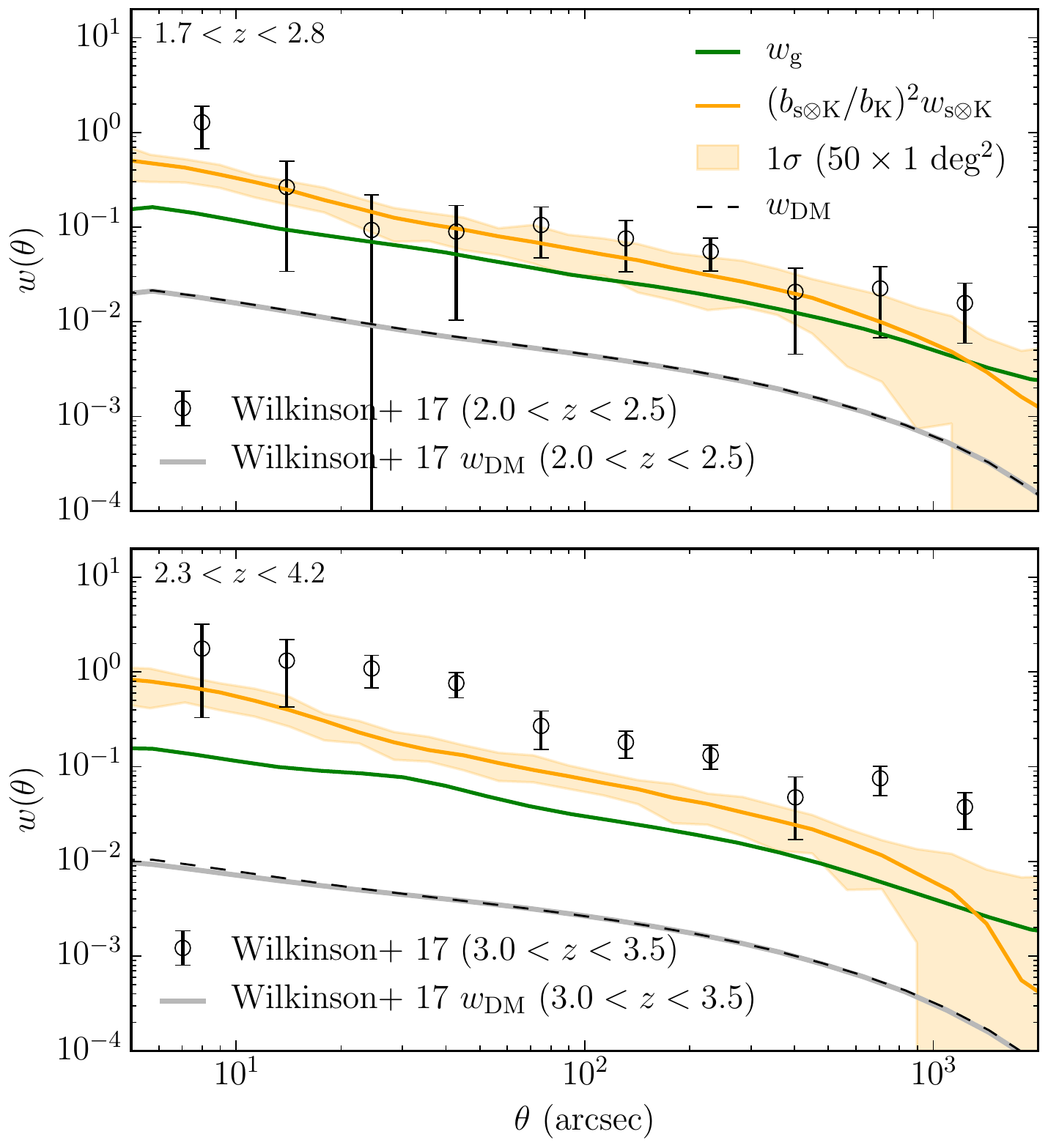}
\caption{Predicted angular correlation functions for the redshift intervals $1.7<z<2.8$ (top panel) and $2.3<z<4.2$ (bottom panel) that correspond to the $2.0<z<2.5$ and $3.0<z<3.5$ intervals in Fig~\ref{fig:wtheta_obs_W16} respectively.  These broader intervals are chosen such that the angular correlation function for dark matter (dashed black line) is in agreement with that used by Wilkinson et al. (solid grey line) for that redshift bin.  All other lines and symbols have the same meaning as in Fig.~\ref{fig:wtheta_obs_W16}.}
\label{fig:wtheta_obs_W16_zbins}
\end{figure}
\begin{figure}
\centering
\includegraphics[width=\columnwidth]{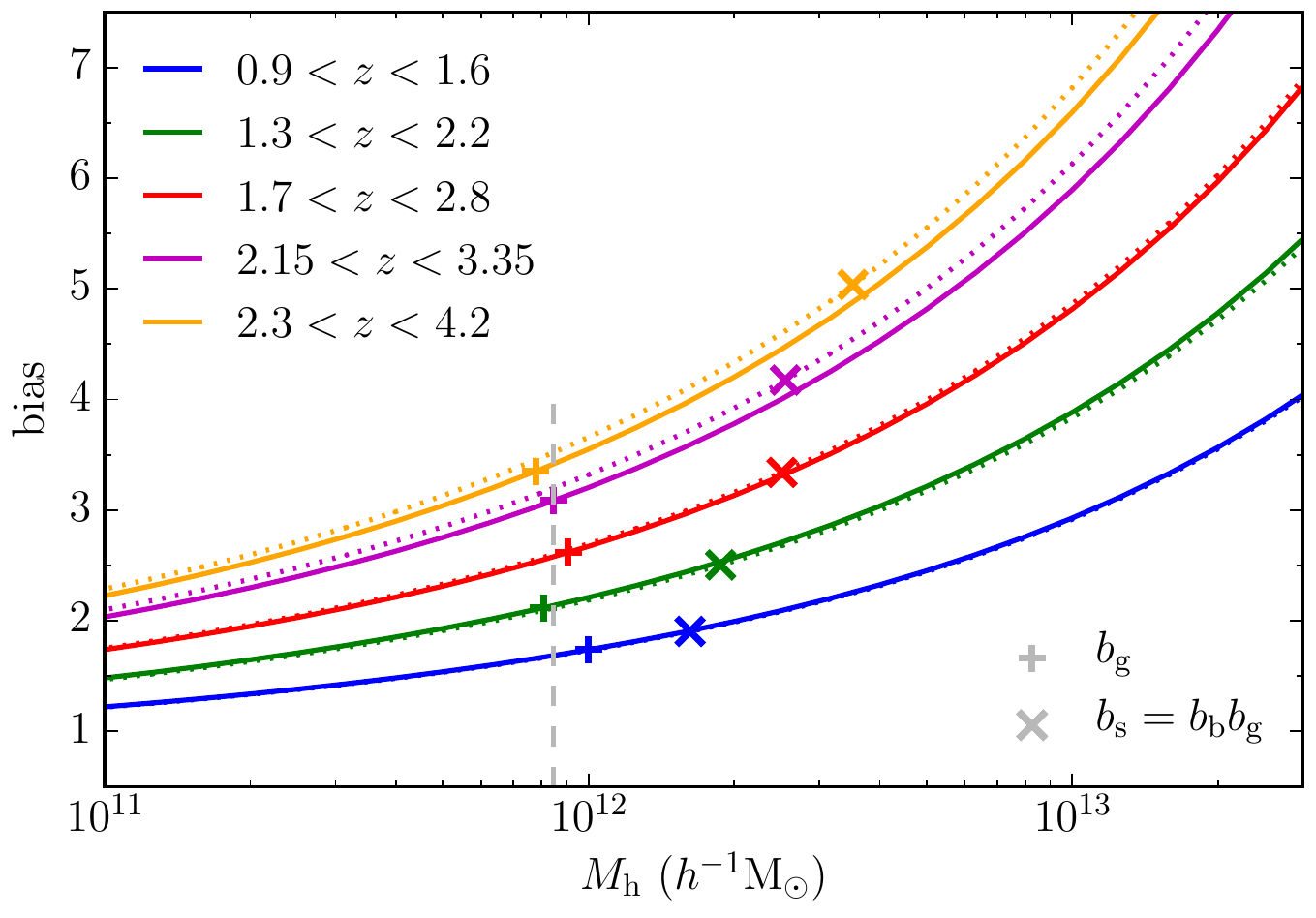}
\caption{Large-scale bias-to-dark matter halo mass relations of Sheth et al (\citeyear{SMT01}) calculated at the median redshift of the interval considered for galaxies (solid lines) and source counterparts (dotted lines).  The different colours are for the redshift intervals indicated in the legend.  Plus signs (crosses) indicate the position on this plane for galaxies (source counterparts) using the biases derived from the corresponding angular correlation functions.  The vertical dashed grey line shows the median inferred halo mass for the galaxies.}
\label{fig:bias_dmh_zbins}
\end{figure}

Given the apparent good agreement between the predicted and observed correlation functions in Fig.~\ref{fig:wtheta_obs_W16}, the cause of the extreme differences in the derived bias values (and subsequent conclusions about the host halo masses) seen in Fig.~\ref{fig:bias_evol_W16} is worthy of further investigation.  As mentioned earlier we attribute this to the width of the photometric redshift probability distributions used for each galaxy by \cite{Wilkinson16}, a necessary consequence of the available photometry.  The redshifts in Wilkinson et al. were mostly obtained from the UDSz ESO Large Programme (ID:180.A-0776; PI: O. Almaini).  The \textsc{EAZY} template-fitting pack \citep{Brammer08} was used to derive a photometric redshift probability distribution for each galaxy through a maximum likelihood analysis.  SMG counterparts were assigned using the OIRTC method \citep{TC16}.  A galaxy in the Wilkinson et al. analysis is able to appear in multiple redshift intervals, weighted by the integral of its probability distribution between the limits of the redshift interval.

A consequence of this is that the effective redshift distributions used for each bin are typically broader than the quoted limits of the bin would suggest, and become broader with increasing redshift as the quality of the photometric redshifts generally degrade i.e. the probability distributions become broader. In Fig.~\ref{fig:Wilkinson16_pzs}, we show the average SMG counterpart redshift distributions for each redshift interval from \cite{Wilkinson16}, these being calculated as the sum of all the individual galaxy photometric redshift probability distributions weighted by their interval between the quoted limits.   

Thus the angular correlation functions for dark matter used by Wilkinson et al. would typically have a lower normalisation than shown in Fig.~\ref{fig:wtheta_obs_W16} (where we used the true redshifts of the galaxies in the simulation and a top-hat redshift window of $\Delta z = 0.5$) as the spatial correlation function of the dark matter, $\xi_{\rm DM}(r,z)$, has effectively been projected over a larger volume.  This explains how the agreement between the angular correlation functions seen in Fig~\ref{fig:wtheta_obs_W16} is consistent with the disagreement in the inferred large-scale bias seen in Fig.~\ref{fig:bias_evol_W16}.  

To mimic the effect of the width of photometric redshift distribution to first order, we increase the width of the redshift intervals we consider (symmetrically in redshift, maintaining the same central redshift) until our dark matter correlation functions have a similar normalisation to those calculated using the redshift distributions of Wilkinson et al. for each bin. These new redshift interval widths are shown in the bottom panel of Fig.~\ref{fig:Wilkinson16_pzs} and listed in column (b) of Table~\ref{table:results}.

We then repeat our analysis using these new top-hat redshift intervals.  We show two examples of this, for the $2.0<z<2.5$ and $3.0<z<3.5$ bins (for which we now use redshift intervals of $1.7<z<2.8$ and $2.3<z<4.2$ respectively), in Fig.~\ref{fig:wtheta_obs_W16_zbins} and list the results for each interval in Table~\ref{table:results}.  Considering a broader redshift distribution brings the large-scale bias values we measure for the simulated sub-mm sources into broad agreement with the values quoted by \cite{Wilkinson16}, apart from the $1.0<z<1.5$ bin where the large-scale bias is overpredicted, and the $3.0<z<3.5$ bin where it is underpredicted. 

Our reasoning for the agreement between the observed and predicted large-scale bias values for sub-mm sources found here is as follows.  As the width of the redshift interval we consider increases, the blending bias also increases.  This is due to the inclusion of more `beam-induced' correlated pairs in the correlation function calculation as is discussed in C16.  However, the intrinsic galaxy large-scale bias remains approximately constant.  Therefore, the increase in blending bias means that the inferred large-scale bias for the sources becomes greater.  

In Fig.~\ref{fig:bias_dmh_zbins} we show the effect this has on the inferred host halo masses as a function of redshift.  We use the large-scale bias-to-halo mass relations of \cite{SMT01} and assume that the objects occupy halos of a single mass at the median redshift of the interval considered, which we calculate using the relevant redshift distributions from Fig~\ref{fig:dndz}.  For the galaxies we find this yields inferred halo masses consistent with those that the galaxies are known to occupy in the model (see Fig.~$5$ of C16) and with no significant redshift evolution over this range.  For the sources however, we observe a mild evolution in halo mass from $\sim4\times10^{12}$~{\hMsol} at $z\sim3$ to $\sim2\times10^{12}$~{\hMsol} at $z\sim1$, due to the blending bias being larger at higher redshift as the redshift interval considered is broader.  Whilst it appears unlikely from this analysis that this effect could account for all of the very strong halo mass `downsizing' found by Wilkinson et al., it is possible that the apparent downsizing trend was amplified by this effect, as the broadening of the redshift intervals with increasing redshift was not considered by Wilkinson et al. when deriving their blending bias factors.                   
  
We conclude that measuring the correlation function for sub-mm sources via an auto- or cross-correlation is affected by blending bias in the same way.  Measuring the cross-correlation using obects within a relatively narrow redshift range is the best way to perform such a measurement, due to the increased statistical significance from the cross-correlation with a more abundant sample and to the reduced blending bias due to the narrower redshift range being investigated.  Such an analysis is performed by \cite{Wilkinson16}.  However, this comes with the important caveat that accurate redshifts for the correct counterpart to the sub-mm source are required, and there are a sufficient number of objects in each redshift bin for the result to be statistically significant.  Alternatively, as is also discussed in C16, a significant targeted follow-up campaign with interferometers such as ALMA would allow the blended sources in the single-dish catalogue to be identified and removed from the clustering analysis, providing a result free from blending bias.  Investigation of the evolution of the SMG clustering with redshift will still require accurate redshifts (at the level that the typical redshift error is expected to be factors of a few smaller than the width of the redshift bin), but this is an issue separate from the blending bias. 

\section{Conclusions}
\label{sec:conclusion}
We study the effect of `blending bias,' the square root of the factor by which the angular correlation function of sources identified through source extraction of confused imaging is boosted relative to that of the underlying galaxy population.  In particular we focus on its applications to the measurement of the clustering of SMGs, an important population of galaxies that exhibit some of the highest inferred star formation rates in the Universe, as this can constrain the dark matter halo masses of these galaxies and thus their subsequent evolution.

To do so we use the galaxy formation model presented in \cite{Lacey16}, which can successfully reproduce the observed number counts of SMGs at $850$~{\mum}, and the methodology of \cite{Cowley15} to create simulated imaging based on the model galaxies.      

We compare our model predictions to the recent analysis of \cite{Wilkinson16}, who cross-correlated a sample of SMGs in the UKIDSS UDS field with a more numerous $K$-band selected sample to derive the large-scale bias and halo masses of SMGs in $5$ redshift intervals from $1.0\lesssim z\lesssim3.5$. 

Importantly we find that the blending bias factors are essentially the same whether the correlation function is derived through an auto- or cross-correlation technique, though they can be reduced by decreasing the width of the redshift interval considered.  This adds weight to the accuracy of the \cite{Wilkinson16} study, which is the first to measure the evolution in the clustering of SMGs. However, our predictions indicate their results may still be affected systematically by blending biases of at least $b_{\rm b}\sim1.1-1.2$, which can lead to the host halo masses being overestimated by a factor of $\sim2$.

We investigate the effect of the redshift intervals considered becoming effectively broader with increasing redshift due to the use of photometric redshift probability distributions, a necessary consequence of the available photometry used in observational studies.  We find that this can result in inferring a spurious halo mass `downsizing' trend, where the halo masses inferred from the clustering at $z\sim3$ are a factor of $\sim2$ greater than those inferred at $z\sim1$.  This is due to the blending bias factors being larger at higher redshift as a result of the broader redshift distributions used.  However, this trend is not as strong as the one observed by \cite{Wilkinson16}. 

Finally, we note that the blending bias values quoted in this work may be somewhat model dependent and caution that further work is required to fully understand the implications of this bias for measurements made from catalogues derived from single-dish imaging surveys at FIR/sub-mm wavelengths.  Additionally, we hope that this effect will be confronted directly with future ALMA observations, which would allow the clustering of an SMG sample to measured free from blending bias.
  
\section*{Acknowledgements}
The authors would like to thank Nuala McCullagh for helpful discussions relating to the evolution of the matter power spectrum, and Violeta Gonzalez-Perez for helpful discussions relating to the evaluation of the Limber equation, for both this work and C16.  WIC and CGL acknowledge financial support and fruitful discussions from the Munich Institute for Astro- and Particle Physics (MIAPP) DFG cluster of excellence `Origin and Structure of the Universe' workshop entitled `The Star Formation History of the Universe', held in Munich in August 2015.  This work was supported by the Science and Technology Facilities Council [ST/K501979/1, ST/L00075X/1].  CMB acknowledges the receipt of a Leverhulme Trust Research Fellowship.  This work used the DiRAC Data Centric system at Durham University, operated by the Institute for Computational Cosmology on behalf of the STFC DiRAC HPC Facility (www.dirac.ac.uk). This equipment was funded by BIS National E-infrastructure capital grant ST/K00042X/1, STFC capital grant ST/H008519/1, and STFC DiRAC Operations grant ST/K003267/1 and Durham University. DiRAC is part of the National E-Infrastructure. 
\bibliographystyle{mnras}
\bibliography{ref} 
\bsp	
\label{lastpage}
\end{document}